\documentclass[twocolumn]{aastex631}

\usepackage{amsmath}
\usepackage{ulem}
\usepackage{xcolor}
\usepackage{amsmath}
\usepackage{cases}
\usepackage[T1]{fontenc}
\usepackage{subfigure}
\usepackage{epsfig}

\shorttitle{Progenitor and Central Engine of short-duration GRB 201006A}
\shortauthors{Tian et al.}

\begin{document}
\title{The Progenitor and Central Engine of short-duration GRB 201006A associated with a coherent radio flash}

\author{Xiao Tian}
\affiliation{Guangxi Key Laboratory for Relativistic Astrophysics, Department of Physics, Guangxi University, Nanning 530004, China; lhj@gxu.edu.cn}

\author[0000-0001-6396-9386]{HouJun L\"{u}} 
\affiliation{Guangxi Key Laboratory for Relativistic Astrophysics, Department of Physics, Guangxi University, Nanning 530004, China; lhj@gxu.edu.cn}

\author{Yong Yuan} 
\affiliation{School of Physics Science And Technology, Wuhan University, No.299 Bayi Road0, Wuhan, 430072 Hubei, China}

\author{Xing Yang} 
\affiliation{Guangxi Key Laboratory for Relativistic Astrophysics, Department of Physics, Guangxi University, Nanning 530004, China; lhj@gxu.edu.cn}

\author{HaoYu Yuan} 
\affiliation{Department of Astronomy, School of Physics, Huazhong University of Science and Technology, Wuhan, 430074, China}

\author{ShuangXi Yi} 
\affiliation{School of Physics and Physical Engineering, Qufu Normal University, Qufu 273165, China}

\author{WenLong Zhang} 
\affiliation{School of Physics and Physical Engineering, Qufu Normal University, Qufu 273165, China}
\affiliation{Key Laboratory of Particle Astrophysics, Institute of High Energy Physics, Chinese Academy of Sciences, Beijing 100049, China}

\author[0000-0002-7044-733X]{EnWei Liang}
\affiliation{Guangxi Key Laboratory for Relativistic Astrophysics, Department of Physics, Guangxi University, Nanning 530004, China; lhj@gxu.edu.cn}

\begin{abstract}
Recently, the detection of a coherent radio flash associated with short-duration GRB 201006A, occurring 76.6 minutes after the burst, has attracted great attention. However, the physical origin of the coherent radio flash remains under debate. By reanalyzing its data observed by Fermi and Swift, we find that an early radio afterglow as the physical origin of the radio flash can be ruled out, but the coherent radio emission seems to be consistent with the hypothesis of a supramassive magnetar as the central engine collapsing into a black hole. Within this scenario, the derived magnetar surface magnetic field ($B_{\rm p}$) and the initial spin period ($P_{\rm 0}$) fall into a reasonable range but require a preferable low value of $\eta_{\rm R} = 10^{-7}$ or $10^{-6}$. Moreover, the calculated low-$\varepsilon$ value and $E_{\rm \gamma,iso}-E_{\rm p}$ correlation of GRB 201006A also supports the progenitor which is from the merger of compact stars. We also discuss the non-detected kilonova emission associated with GRB 201006A, and then compare with its upper limits of optical observations. 

\end{abstract}

\section{Introduction}
The mergers of compact binary star systems, comprising either two neutron stars (NS-NS; \citealt{1986ApJ...308L..43P, 1989Natur.340..126E}) or an NS and a black hole (NS-BH; \citealt{1991AcA....41..257P}), are commonly believed to be the progenitor of short-duration gamma-ray bursts (SGRBs). The simultaneous detection of the gravitational-wave event GW170817 and its electromagnetic counterpart (SGRB 170817A and AT2017agfo) provides “smoking gun” evidence to support this hypothesis \citep{Abbott2017a, Abbott2017b, 2017ApJ...848L..14G, 2017ApJ...848L..15S, 2018NatCo...9..447Z}, and proves that at least some SGRBs do indeed originate from NS-NS mergers. Nevertheless, the nature of the central engine of SGRB remains under debate. Two leading models of the central engine are extensively discussed. One is a hyperaccreting BH \citep{1999ApJ...518..356P, 2013ApJ...765..125L, 2017NewAR..79....1L}; the other is a rapidly spinning, strongly magnetized NS called millisecond magnetar, which powers the outflow of GRBs through its rotation energy \citep{1992Natur.357..472U, 1994MNRAS.270..480T, Dai1998a, Dai1998b, 2001ApJ...552L..35Z, 2011MNRAS.413.2031M, 2012MNRAS.419.1537B, 2014ApJ...785...74L, 2015ApJ...805...89L}.

For NS-BH merger, the remnant must be a BH surrounded by an accretion disk. Within the scenario of NS-NS mergers, four possible outcomes of the merger that are dependent on the nascent NS mass ($M_p$) and its unknown equation of state are expected \citep{2014PhRvD..89d7302L, 2015ApJ...805...89L, 2016PhRvD..94h3010L, 2016PhRvD..93d4065G}. One possibility is a BH when the $M_p$ is much greater than the maximum nonrotating mass ($M_{\rm TOV}$; \citealt{2003MNRAS.345.1077R, 2011ApJ...732L...6R, 2014MNRAS.441.2433R}). Another possible remnant is a magnetar. Depending on the relationship between $M_p$, $M_{\rm TOV}$, and the maximum gravitational mass ($M_{\rm max}$), three possible remnants may be formed for the post-merger evolution of magnetars: (1) a hypermassive NS, which is supported exclusively by its differential rotation, can survive hundreds of milliseconds \citep{2000A&A...360..171R, 2010MNRAS.406.2650M}; (2) a supramassive NS, which can be supported by its rigid rotation, can survive from tens of seconds to thousands before collapsing into a BH \citep{2010MNRAS.409..531R, 2013ApJ...763L..22Z, 2015ApJ...805...89L, 2016PhRvD..93d4065G, 2017ApJ...835..181L}; (3) a stable NS with a much longer lifetime \citep{2006Sci...311.1127D, 2015PhR...561....1K, 2018MNRAS.480.4402L}. On the other hand, an optical/infrared transient with near-isotropic (called kilonova) can be generated from the ejected materials and powered by radioactive decay from r-process after binary NS mergers \citep{1998ApJ...507L..59L, 2010MNRAS.406.2650M, 2013ApJ...774L..23B, 2013ApJ...776L..40Y, 2015NatCo...6.7323Y, 2016NatCo...712898J, 2021ApJ...912...14Y,  2022ApJ...931L..23L, 2023Univ....9..245T}. A small fraction of SGRBs reported to be associated with kilonova candidates are already confirmed from the observations (see \citealt{2019LRR....23....1M} for a review).

From the observational point of view, the X-ray ``internal plateaus'' with rapid decay at the end of the plateaus in some short GRBs \citep{2010MNRAS.409..531R, 2013MNRAS.430.1061R, 2015ApJ...805...89L, 2017ApJ...835..181L}, are difficult to interpret within the framework of a BH central engine but are consistent with a supramassive magnetar as the central engine \citep{2006Sci...311.1127D, 2010MNRAS.409..531R, 2013MNRAS.431.1745G}. The rapid decay followed the plateau is suggested to originate from a supramassive magnetar collapsing into a BH \citep{2013ApJ...763L..22Z, 2015ApJ...805...89L, 2017ApJ...835..181L, 2015PhR...561....1K}. On the other hand, the magnetar model is not the only way to explain X-ray plateau emission; for example, the structured jets viewed from off-axis can also explain some X-ray plateau emissions \citep{2020MNRAS.492.2847B}. More interestingly, \cite{2014A&A...562A.137F} first proposed that a possible radio emission (i.e., fast radio burst (FRB)) can be produced when a spinning supramassive NS loses centrifugal support and collapses into a BH. \cite{2014ApJ...780L..21Z} proposed that this radio emission is possible associated with GRBs. Within this picture, the radio emission would be physically connected to the internal plateau in SGRBs. \cite{2012ApJ...757...38B} reported that an upper limit of prompt radio emission (i.e., FRB-like event) seems to be consistent with a long-duration GRB at the end of plateau emission, but can not be confirmed.

Recently, \cite{2024MNRAS.tmp.2177R} claimed that the detection of a coherent radio flash with $5.6\sigma$ confidence is associated with the SGRB 201006A. The short radio flash of 144 MHz is detected 76.6 minutes after the burst trigger time. They proposed that the coherent radio flash is powered by the collapse of a long-lasting supramassive NS as the central engine into a BH, and the surviving supramassive NS originates from a binary NS merger \citep{2024MNRAS.tmp.2177R}. However, \cite{2024ApJ...973L..20S} proposed that the coherent radio emission was powered far from a BH central engine via synchrotron maser or magnetic reconnection in the jet. In any case, the central engine of SGRB 201006A associated with the coherent radio flash remains under debate. If SGRB 201006A is indeed associated with the coherent radio flash, then several questions emerge. What is the progenitor of GRB 201006A? Could the coherent radio flash be an early radio afterglow resulting from the interaction between the jet and the interstellar medium? What are the physical parameters of a supramassive NS as the central engine when we assume that the radio flash originates from the collapse of a supramassive NS into a BH?

In this paper, we systematically analyze the observational data of the prompt emission (in Section 2). Then, we identify the progenitor of GRB 201006A by comparing it with the other type I and type II GRBs in Section 3. In section 4, we attempt to determine whether the coherent radio flash could be an early radio afterglow (i.e., forward shock (FS) or reverse shock (RS) of external shock model) or a supramassive NS as the central engine collapsing into a BH. The conclusions are drawn in Section 5 with some discussion. Throughout the paper, the convention $Q = 10^{x}Q_{x}$ in cgs units and a concordance cosmology with parameters $\Omega_{\rm M} = 0.30$, $\Omega_{\rm \Lambda} = 0.70$ and $H_{0} = 70 ~\rm km ~\rm s^{-1} ~Mpc^{-1}$ are adopted. 

\section{The Observations and Data Analysis}
\subsection{Fermi Data Reduction}
At 01:17:52.27 UT on 2020 October 6, the Fermi Gamma-Ray Burst Monitor (GBM) triggered and located GRB 201006A \citep{2020GCN.28564....1H}. Fermi/GBM is configured with 12 sodium iodide (Na I) and two bismuth germanate scintillation detectors with energy detection ranging from 8keV to 40MeV \citep{2009ApJ...702..791M}. The corresponding time-tagged event data were downloaded from the Fermi/GBM public data website\footnote{https://heasarc.gsfc.nasa.gov/FTP/fermi/data/gbm/daily/}. Note that more detailed information about light curve and spectral procedure can be found in \cite{2016ApJ...816...72Z}. Here the light curves of n8, n9 and b1 detectors are shown in Figure 1. We notice that the duration ($T_{90}$) of GRB 201006A  in the energy band 50-300 keV is reported to be 1.7 s \citep{2020GCN.28564....1H}.

We also extract the time-averaged spectrum of GRB 201006A within the time interval from $T_{0}$-0.19 s to $T_{0}$+0.83 s, where $T_{0}$ is the trigger time. The background spectra are extracted by selecting two time intervals before and after the burst. We model the background using empirical function \citep{2011ApJ...730..141Z} and invoke XSPEC to fit the spectra. A variety of spectral models can be employed to test the spectral fitting, including power law (PL), cutoff PL (CPL), blackbody (BB) and Band function (Band), or even the union of any two models. Subsequently, the Bayesian information criteria (BIC)\footnote{The BIC is a criterion for model selection in a limited set of models. The model with the lowest BIC value is the best. The BIC values of different models are presented as 477 (Band), 358 (BB), 293 (CPL), 505 (PL), 369 (Band+BB), 517 (BB+PL), and 304 (BB+CPL).} is employed, revealing that the CPL model exhibits superior goodness of the fits and emerges as the best choice for describing the observed data. Figure 2 shows the fitting results of the CPL model, containing the photon spectrum and parameter constraints of the fit. One has peak energy $E_{\rm p} = 103\pm58\rm~ keV$ and a lower energy spectral index $\alpha = -1.03\pm0.30$. According to the spectral analyses, the estimated fluence within $1-10^{4}~\rm keV$ energy band during the time interval is $3.12^{+0.58}_{-0.53} \times 10^{-7} \rm erg~ cm^{-2}$. 

\subsection{Swift Data Reduction}
The Burst Alert Telescope (BAT) of Swift also triggered on 2020 October 6 \citep{2020GCN.28567....1B}. We obtained public BAT data from the Swift archive\footnote{https://www.swift.ac.uk/archive/selectseq.php?source=obs\&tid=957271} and used the standard HEASOFT (v6.28) tools to process these data. The light curves in different energy bands are extracted by adopting batbinevt \citep{2008ApJS..175..179S} with fixed 128 ms time bin, and it consists of a single hard spike shown in Figure 1. The official website of Swift provides a value of $0.49\pm0.09$ s for the $T_{\rm 90}$ in the energy range from 15 to 350 keV. 

The Swift X-ray telescope (XRT) began observing the field
at 83.9 s after the BAT trigger. We made use of the public data from the Swift archive11 \footnote{https://www.swift.ac.uk/xrt\_curves/00998907/} \citep{2020GCN.28568....1G, 2020GCN.28562....1G}. The X-ray light curve seems to be a PL decay with a decay index of 0.99 (see Figure 6). The Chandra X-ray Observatory also starts to observe GRB 201006A at 3.98 days post-trigger \citep{2020GCN.28598....1R}, and does not detect an X-ray source within the enhanced Swift-XRT position but obtain a 3$\sigma$ limit of $3.6\times 10^{-4} \rm cts/s$ in the 0.5-8 keV energy range.

The Ultra-Violet Optical Telescope (UVOT) began observations of GRB 201006A at 88 seconds after BAT trigger, with no optical afterglow consistent with the XRT position \citep{2020GCN.28565....1M}. Moreover, the GROWTH-India Telescope (GIT), Lowell Discovery Telescope (LDT), and MITSuME are also follow-up to observe this source but do not find any source in the stacked image, only an upper limit of $r>20.09$ mag at 17.25 hours, $R>19.4$ mag at 12.2 hours, $i>23.8$ mag at 1.44 days after trigger, respectively \citep{2020GCN.28573....1K, 2020GCN.28571....1I, 2020GCN.28572....1D}.

\subsection{Estimated the redshift of GRB 201006A}
Since no optical counterpart was detected, along with no identified host galaxy within the near-infrared depth range of the X-ray counterpart location for GRB 201006A, one has to estimate the distance of GRB 201006A via dispersion of radio emission. \cite{2024MNRAS.tmp.2177R} claimed that the 3.8$\sigma$ detection significance of radio flash associated with GRB 201006A corresponds to a dispersion measure value of 740-800 pc $\rm cm^{-3}$ when the intrinsic duration is about 5 s. They adopted the Macquart correlation \citep{2022MNRAS.509.4775J} to estimate a redshift of GRB 201006A at $z=0.58\pm0.06$ by taking into account the DM range and the contributions of the Galactic to the dispersion measure along the line of sight to GRB 201006A. In this paper, we adopt the redshift $z=0.58$ to do the calculations.

\section{Progenitor of GRB 201006A}
Phenomenally, GRBs can be classified into two categories (``long-soft'' versus ``short-hard'') based on the duration ($T_{90}$) and spectral hardness of the prompt emission, and the division line is at the duration $T_{90} \sim 2$ s \citep{1993ApJ...413L.101K, 2013ApJ...764..179B}. Several lines of observational evidence show that some LGRBs are associated with core-collapse supernovae (SNe; e.g.,\citealt{1993A&AS...97..205W, 1998Natur.395..670G, 2003ApJ...591L..17S, 2004ApJ...609L...5M, 2006ApJ...645L..21M, 2006Natur.442.1011P}), and suggest that LGRBs may originate from the death of massive stars (type II) \citep{2006Natur.444.1010Z, 2010ApJ...725.1965L, 2019LRR....23....1M}. In contrast, short GRBs are generally attributed to the merger of two compact stars (type I), due to the fact that some SGRBs are potentially associated with kilonova and GW radiation rather than SNe and typically occur in regions of the host galaxy with little star formation \citep{2006Natur.444.1010Z, 2007ApJ...655L..25Z, 2010ApJ...725.1965L, 2013ApJ...774L..23B, 2016NatCo...712898J, Abbott2017b, 2017ApJ...835..181L, 2019ApJ...870L..15L, 2019LRR....23....1M, 2023Univ....9..245T}. However, the measurement of $T_{90}$ is energy and instrument dependent \citep{2013ApJ...763...15Q}. Some short-duration GRBs may be from death of massive stars (e.g., GRB 200826A; \citealt{2021NatAs...5..917A, 2021NatAs...5..911Z, 2022ApJ...932....1R}), and some long-duration GRBs are from mergers of two compact stars (e.g., GRB 060614, \citealt{2006Natur.444.1044G, 2015NatCo...6.7323Y}; GRB 211211A, \citealt{2022Natur.612..223R, 2022Natur.612..228T, 2022Natur.612..232Y, 2023ApJ...943..146C, 2023NatAs...7...67G}; GRB 211227A, \citealt{2022ApJ...931L..23L, 2023A&A...678A.142F}; GRB 230307A, \citealt{2023ApJ...954L..29D, 2023arXiv230705689S, 2024Natur.626..737L, 2024Natur.626..742Y, 2024ApJ...962L..27D, 2024ApJ...963L..26Z}).

\cite{2010ApJ...725.1965L} proposed a new phenomenological classification method for GRBs. They introduced a new parameter $\varepsilon$ which is defined as
\begin{equation}\label{eq:1}
\varepsilon = E_{\rm \gamma,iso,52}/E^{\rm 5/3}_{\rm p,z,2},
\end{equation}
where $E_{\rm \gamma,iso,52}$ is the isotropic burst energy and $E_{\rm p,z,2} = E_{\rm p}\rm (1+z)/100keV$ is the rest-frame peak energy. They found that the $\varepsilon$ has a clear bimodal distribution (high- and low- $\varepsilon$ regions) with a division line at $\varepsilon \sim 0.03$, and the low- and high-$\varepsilon$ regions correspond to the mergers of two compact stars (type I) and death of massive stars (type II), respectively. The short-duration GRB 201006A with estimated redshift $z = 0.58\pm0.06$ has an isotropic energy $E_{\rm \gamma,iso} = 2.76^{+0.51}_{-0.47} \times 10^{50} \rm erg$ within $1-10^{4}~\rm keV$ and peak energy $E_{\rm p} = 103\pm58\rm~ keV$, respectively. By adopting the above method, one can calculate $\varepsilon \sim 0.012$, and it is located in the low-$\varepsilon$ region. The low-$\varepsilon$ value of GRB 201006A suggests that it is consistent with type I population from the compact star mergers (see Figure 3).

Observationally, on the other hand, some empirical correlations among several observed quantities have been claimed \citep{2009ApJ...703.1696Z}, such as Amati relation $E_{\rm \gamma,iso}-E_{\rm p}$ \citep{2002A&A...390...81A}. It is found that a majority of long-duration GRBs (type II) exhibit a positive correlation as $E_{\rm p}$ ($E_{\rm p, z} \propto E^{1/2}_{\rm \gamma,iso}$), even the dispersion of the correlation is large, and outliers do exist \citep{2009ApJ...703.1696Z}. However, 
the Amati relation of most short-duration GRBs (type I) are inconsistent with that of long-duration GRBs, and it seems to be a little bit shallower for power index compared with that of long GRBs \citep{2009ApJ...703.1696Z}. In order to test whether GRB 201006A obeys the empirical correlation of $E_{\rm \gamma,iso}-E_{\rm p}$, we plot GRB 201006A in the $E_{\rm p}-E_{\rm \gamma,iso}$ diagram and compare with other type I and type II GBRs (see Figure 3). It is found that GRB 201006A deviates from the correlation of type II GRBs but seems to be closer to type I GRBs. In any case, the $\varepsilon$ method and Amati relation indicate that the progenitor of GRB 201006A should originate from the merger of compact stars, e.g., NS-NS or NS-BH. Together with the observed coherent radio emission associated with GRB 201006A, the merger of NS-NS is likely to be a potential candidate of the GRB 201006A progenitor.

\section{Possible physical origin of radio flash associated with GRB 201006A}
In this section, we present more details of two possible physical origins of the radio flash associated with GRB 201006A, e.g., afterglow origin, and the supramassive NS central engine collapsing into a BH. Also, we compare with the observations and discuss the possible kilonova emission within the hypothesis of NS-NS merger.  

\subsection{Origin of forward/reverse shock of afterglow?}
Recently, \cite{2024MNRAS.tmp.2177R} claimed to detect a coherent radio flash that is associated with GRB 201006A with $5.6\sigma$ confidence. The short radio flash of 144 MHz with duration of 5 s was detected at 76.6 minutes (in observer frame) after the burst trigger time, and the peak flux density of the radio flash was $49\pm27$ mJy with a redshift of $0.58 \pm 0.06$. Within the framework of fireball model of GRB, the interaction between relativistic jets and the surrounding medium via synchrotron radiation can give rise to multiwavelength afterglow (X-ray, optical and radio) emissions of GRBs \citep{1997ApJ...476..232M, 1998ApJ...497L..17S, 2013NewAR..57..141G, 2014ApJ...792L..21Y}. One question is whether the coherent radio flash is an early radio afterglow of SGRB 201006A from an external shock. In this section, we adopt the standard external shock with synchrotron emission afterglow model to calculate the flux of the possible early radio emission.

By considering a standard fireball, the afterglow is mainly determined by the initial Lorentz factor ($\Gamma$) and total kinetic energy ($E_{\rm K, iso}$). The interaction between the relativistic jet and the ambient medium can produce a pair of shocks (forward and reverse) to propagate into the surrounding medium and the ejecta, respectively. Following our previous work in \cite{2014ApJ...792L..21Y}, the time of crossing the shell by RS is the deceleration time, which can be expressed as 
\begin{equation}\label{eq:2}
t_{\rm dec} \sim \frac{l(1+\rm z)}{2c\Gamma^{8/3}},
\end{equation}
where $l=(3E_{\rm K, iso}/4\pi nm_p c^2)$ is the Sedov length. The initial total kinetic energy $E_{\rm K, iso} =E_{\rm \gamma,iso} (1-\eta_{\gamma})/\eta_{\gamma}$, where $\eta_{\gamma}$ is the radiation efficiency of GRBs. By adopting $\eta_{\gamma} \sim 0.2$, one has $E_{\rm K, iso} \sim 10^{51} \rm~ erg$ for GRB 201006A owing to $E_{\rm \gamma,iso} = 2.76^{+0.51}_{-0.47} \times 10^{50} \rm~ erg$ in $\gamma-$ray emission. On the other hand, we also adopt the typical ambient medium density value of short GRBs ($n=10^{-2}$~$\rm cm^{-3}$) and $\Gamma = 100$ to calculate the afterglow flux, and one has $t_{\rm dec} \sim 307~\rm s$. The evolution of the light curves of FS and RS is related to three characteristic frequencies: minimum synchrotron frequency ($\nu_{\rm m}$), cooling frequency ($\nu_{\rm c}$), and self-absorption frequency ($\nu_{\rm a}$). $F_{\rm \nu,max}$ is the peak flux of the spectrum.

(1) In the case of FS, based on the standard afterglow model \citep{1998ApJ...497L..17S, 2013NewAR..57..141G, 2013ApJ...776..120Y, 2023RAA....23k5010D}, at the deceleration time $t_{\rm dec}$, the characteristic parameters of frequencies for FS emission can be expressed as:
\begin{equation}\label{eq:3}
\nu_{\rm m,FS} = 4.1 \times 10^{15} \epsilon^{1/2}_{\rm B,f,-2} \epsilon^{2}_{\rm e,-1} n^{1/2}_{-2} \Gamma^4_2~ \rm Hz,
\end{equation}
\begin{equation}\label{eq:4}
\nu_{\rm c,FS} = 7.5 \times 10^{18} \epsilon^{-3/2}_{\rm B,f,-2} n^{-5/6}_{-2} \Gamma^{4/3}_{2} E^{-2/3}_{51}~ \rm Hz,
\end{equation}
\begin{equation}\label{eq:5}
\nu_{\rm a,FS} = 2.9 \times 10^{8} \epsilon^{1/5}_{\rm B,f,-2} \epsilon^{-1}_{\rm e,-1} n^{3/5}_{-2} E^{1/5}_{51} ~ \rm Hz,
\end{equation}
\begin{equation}\label{eq:6}
F_{\rm \nu,max,FS} = 7.8 \times 10^{-5} \epsilon^{1/2}_{\rm B,f,-2} n^{1/2}_{-2} E_{51} D^{-2}_{\rm L,28}~ \rm Jy.
\end{equation}
Here, we adopt the shock microphysics parameters $\epsilon_{\rm e} = 0.1$, $\epsilon_{\rm B, f} = 0.01$, and the electron injection spectral index is $p = 2.5$. These four parameters of FS before and after the crossing time can be written as \citep{1997ApJ...476..232M, 1998ApJ...497L..17S, 2014ApJ...792L..21Y}:

(a) $t < t_{\rm dec}$, 
\begin{equation}\label{eq:7}
\nu_{\rm a, FS} \propto t^{\frac{3}{5}}, \nu_{\rm m, FS} \propto t^0, \nu_{\rm c, FS} \propto t^{-2}, F_{\rm \nu, max, FS} \propto t^{3},
\end{equation}

(b) $t > t_{\rm dec}$,
\begin{equation}\label{eq:8}
\nu_{\rm a, FS} \propto t^0, \nu_{\rm m, FS} \propto t^{-\frac{3}{2}}, \nu_{\rm c, FS} \propto t^{-\frac{1}{2}}, F_{\rm \nu, max, FS} \propto t^{0}.
\end{equation}
For non-relativistic phase, it would be $\Gamma - 1 = 1$, where $\Gamma \sim (3E_{\rm K, iso}/32 \pi n m_p c^5 t^3)^{1/8}$. After this transition time (the connection time between relativistic phase and nonrelativistic phase), the parameters of FS emission should be modified as
\begin{equation}\label{eq:9}
\nu_{\rm a, FS} \propto t^\frac{6}{5}, \nu_{\rm m, FS} \propto t^{-3}, \nu_{\rm c, FS} \propto t^{-\frac{1}{5}}, F_{\rm \nu, max, FS} \propto t^{\frac{3}{5}}.
\end{equation}

(2) In the case of RS, the four parameters ($\nu_{\rm m, RS}$, $\nu_{\rm c, RS}$, $\nu_{\rm a, RS}$, $F_{\rm \nu, max, RS}$) can be expressed as  
\begin{equation}\label{eq:10}
\nu_{\rm m, RS} = 1.3 \times 10^{12} \epsilon^{1/2}_{\rm B,r,-1} \epsilon^{2}_{\rm e,-1} n^{1/2}_{-2} \Gamma^2_2~ \rm Hz,
\end{equation}
\begin{equation}\label{eq:11}
\nu_{\rm c, RS} = 2.4 \times 10^{17} \epsilon^{-3/2}_{\rm B,r,-1} n^{-5/6}_{-2} \Gamma^{4/3}_2 E^{-2/3}_{51}~ \rm Hz,
\end{equation}
\begin{equation}\label{eq:12}
\nu_{\rm a, RS} = 2.9 \times 10^{11} \epsilon^{1/5}_{\rm B,r,-1} \epsilon^{-1}_{\rm e,-1} n^{3/5}_{-2} \Gamma^{8/5}_2 E^{1/5}_{51}~ \rm Hz,
\end{equation}
\begin{equation}\label{eq:13}
F_{\rm \nu, max, RS} = 2.5 \times 10^{-2} \epsilon^{1/2}_{\rm B,r,-1} n^{1/2}_{-2} \Gamma^2 E_{51} D^{-2}_{\rm L,28} ~ \rm Jy.
\end{equation}
The evolution of these parameters for RS emission is shown as follows.

(a) $t < t_{\rm dec}$,
\begin{equation}\label{eq:14}
\nu_{\rm a, RS} \propto t^{-\frac{33}{10}}, \nu_{\rm m, RS} \propto t^6, \nu_{\rm c, RS} \propto t^{-2}, F_{\rm \nu,max,RS} \propto t^{\frac{3}{2}},
\end{equation}

(b) $t > t_{\rm dec}$,
\begin{equation}\label{eq:15}
\nu_{\rm a, RS} \propto t^{-\frac{102}{175}}, \nu_{\rm m, RS} \propto t^{-\frac{54}{35}}, \nu_{\rm c, RS} \propto t^{{-\frac{54}{35}}}, F_{\rm \nu,max,RS} \propto t^{-\frac{34}{35}}.
\end{equation}

If GRB 201006A originated from the merger of compact stars, the circumburst medium density should be less than $10^{-2}$~$\rm cm^{-3}$ \citep{2013NewAR..57..141G, 2016SSRv..202....3Z}.
Figure 4 shows the numerical calculation of FS and RS afterglow light curves with different parameters (e.g., n, $\epsilon_{\rm B,f}$, and $\epsilon_{\rm B,r}$) of GRB 201006A in the 144 MHz radio afterglow band but fixed $\epsilon_{\rm e} = 0.1$, $p = 2.5$ and $\Gamma = 100$. Note that we allow the microphysics parameter $\epsilon_{\rm B,r}$ to be higher than $\epsilon_{\rm B,f}$ because the outflow is likely magnetized \citep{2003ApJ...595..950Z}. 
In the left panel of Figure 4, we fixed $\epsilon_{\rm B,f} = 0.01$ and $\epsilon_{\rm B,r} = 0.64$ \citep{2014ApJ...792L..21Y} and then plotted the light curves of RS and FS by adopting a variable circumburst medium density, i.e., $n = 10^{-2}$~$\rm cm^{-3}$, $n = 10^{-4}$~$\rm cm^{-3}$, and $n = 10^{-6}$~$\rm cm^{-3}$. In the right panel of Figure 4, we fixed $n=10^{-2}$~$\rm cm^{-3}$ and then adopted different parameters of shock microphysics, i.e., $\epsilon_{\rm B,f} = 10^{-2}, 10^{-3}, 10^{-4}$ for FS model and $\epsilon_{\rm B,r} = 0.1, 0.64, 0.01$ for RS model. It is found that the brightness of FS and RS emissions strongly depends on the values of selected $n$ (left panel of Figure 4), while the varying values of $\epsilon_{\rm B}$ seem to affect a little bit FS and RS light curves in the early time. Moreover, we also compare the observed radio flash with the theoretical calculations of external shock model (e.g., FS and RS models) and find that the flux of the observed radio flash at 76.6 minutes is still much higher than that of any afterglow model. This suggests that an early radio afterglow as the physical origin of the radio flash associated with GRB 201006A can be ruled out.

\subsection{Originated from the collapse of a supramassive magnetar into a black hole?}
A number of previous studies suggested that the X-ray internal plateau following the extremely steep decay phase of SGRBs can be consistent with the collapse of a supramassive NS into a BH \citep{2010MNRAS.409..531R, 2013ApJ...763L..22Z, 2015ApJ...805...89L, 2017ApJ...835..181L}. Meanwhile, the coherent radio emission is expected to be emitted during the collapse \citep{2012ApJ...757...38B, 2014ApJ...780L..21Z}. Within this picture, coherent radio emission and X-ray plateau emission would be simultaneously produced after GRB prompt emission, but this has not been observed so far. \cite{2024MNRAS.tmp.2177R} proposed that the coherent radio flash associated with GRB 201006A is powered by the collapse of a long-lasting supramassive NS as the central engine into a BH, but we do not observe a clear plateau emission in the X-ray afterglow of SGRB 201006A. In this section, we try to constrain the physical parameters of a supramassive NS and compare with observations and other type I GRBs.

Based on the method in \cite{2001ApJ...552L..35Z}, the total rotation energy of a magnetar is 
\begin{equation}\label{eq:16}
E_{\rm rot} = \frac{1}{2} I \Omega_{0} \approx 2 \times 10^{52}~\rm erg~ M_{1.4} R^2_{6} P^{-2}_{0,-3},
\end{equation}
where I, $\Omega_{0}$, R are the moment of inertia, initial angular frequency, and radius of the NS, respectively. $M_{1.4} = M/1.4 M_{\rm \odot}$ is the mass of the NS. In general, the magnetar can lose its rotational energy via both electromagnetic ($L_{\rm EM}$) and gravitational wave ($L_{\rm GW}$) radiations \citep{2001ApJ...552L..35Z, 2013PhRvD..88f7304F, 2016MNRAS.458.1660L}. Here, we assume that the rotation energy loss is dominated by dipole radiation and ignore the contribution from energy loss of gravitational wave. The characteristic spin-down timescale ($\tau$) and spin-down luminosity ($L_0$) can be written as 
\begin{equation}\label{eq:17}
\tau = 2.05 \times 10^{3}~ I_{45}B^{-2}_{p,15}P^{2}_{0,-3}R^{-6}_{6}~s,
\end{equation}
\begin{equation}\label{eq:18}
L_{0} = 1.0 \times 10^{49}~ B^2_{p,15}P^{-4}_{0,-3}R^6_{6}~\rm erg~s^{-1},
\end{equation}
where $B_{p}$ and $P_0$ correspond to the surface polar cap magnetic field and initial spin period of magnetar, respectively. In our calculations, we adopt $L_{\rm EM} \approx L_{0}$. By considering radiatio efficiency ($\eta_{\rm X}$) in the X-ray band, the observed X-ray plateau luminosity can be written as
\begin{equation}\label{eq:19}
L_{\rm X} = \eta_{\rm X} L_{0}.
\end{equation} 
The X-ray radiation efficiency $\eta_{\rm X}$ strongly depends on the injected luminosity $L_{\rm EM}$, and a larger injection luminosity corresponds to a higher radiation efficiency \citep{2019ApJ...878...62X}. From observational point of view, several solid magnetar cases claimed that the efficiency from spin-down luminosity to X-ray emission is as low as $10^{-2}$ or $10^{-3}$, such as for X-ray transient CDF-S XT2 \citep{2019Natur.568..198X, 2019ApJ...878...62X} and GRB 230307A \citep{2023arXiv230705689S}. In this paper, we adopt a constant $\eta_{\rm X}$, fixed as $\eta_{\rm X} = 10^{-3}$. 

If the coherent radio emission is indeed from the collapse of a supramsssive magnetar into a BH, \cite{2014ApJ...780L..21Z} proposed that a supramassive magnetar formed by the merger of two NSs may collapse into a BH within $10^{3}-10^{4}$ s, and magnetic reconnection of the magnetosphere during collapse will produce a short burst of coherent radio emission. If this is the case, the total magnetosphere energy ($E_{\rm B,iso}$) released via this magnetic reconnection process is expected to be 
\begin{eqnarray}\label{eq:20}
E_{\rm B, iso} &\approx& \int^{R_{\rm LC}}_{R}4\pi r^2
\frac{B^2_p}{8\pi}\left(\frac{r}{R}\right)^{-6} dr \nonumber \\& \approx &
(1/6)B^2_p R^3 \approx (1.7\times 10^{47})\, B^2_{p, 15} R^3_{6}~ {\rm erg},
\end{eqnarray}
where $R_{\rm LC} \gg R$ is the light cylinder radius. One can define the conversion efficiency ($\eta_{\rm R} = E_{\rm R,iso}/E_{\rm B, iso}$), which is the fraction of magnetosphere energy converted to coherent radio emission. Here, $E_{\rm R,iso}$ is the isotropic energy of radio flash, and it can be calculated as
\begin{equation}\label{eq:21}
E_{\rm R,iso} = 4 \pi D^{2}_{\rm L}f/(1+z),
\end{equation} 
where $f$ is the observed fluence of radio emission.

By applying the above physical process to GRB 201006A, the observed fluence of radio flash associated with GRB 201006A is $245\pm135$ Jy ms \citep{2024MNRAS.tmp.2177R}, and one can calculate $E_{\rm R,iso} = (3.06\pm1.69) \times 10^{41}~\rm erg$. In theory, the parameter of $\eta_{\rm R}$ is poor unknown, but from an observational point of view, \cite{2023NatAs...7..579M} calculated the solid case for FRB 190425 by invoking a blitzar model and determined an efficiency as low as $2\times 10^{-6}$. Based on Eq.(\ref{eq:20}), we adopt the variable values of $\eta_{\rm R} = 10^{-4}, 10^{-5}, 10^{-6}, 10^{-7}$ to estimate the $B_{\rm p}$ for fixed $R=10$ km, and the results are shown in Table 1. Since we do not observe the X-ray plateau emission of GRB 201006A, the spin-down timescale ($\tau$) cannot be confirmed. However, one can present the lower limit of $\tau$ as the time of coherent radio emission, which is also the time of supramassive magnetar collapse into a BH ($t_{\rm col}$), namely, $\tau \geq t_{\rm col}\approx 49$ minutes (in the rest-frame; adopting $z = 0.58$). By combining with Eqs.(\ref{eq:17}) and (\ref{eq:18}), one can estimate the lower limit of $P_{0}$ for varying $\eta_{\rm R}$, and the results are also shown in Table 1. Figure 5 shows the distribution of $B_{p}$ and $P_{0}$ of GRB 201006A, and compares with other short GRBs taken from \cite{2015ApJ...805...89L}. We find that $B_{p}$ of GRB 201006A for different $\eta_{\rm R}$ seems to be one order less than that of other short GRBs. It is a natural explanation with the long-lasting X-ray plateau emission and lower spin-down plateau luminosity. It is worth noting that the initial spin period $P_{0}$ is below 0.96 ms, which is the breakup spin period limit for a NS \citep{2004Sci...304..536L} for $\eta_{\rm R} = 10^{-4}, 10^{-5}$, and it suggests that $\eta_{\rm R}$ seems to prefer lower values, e.g., $\eta_{\rm R} = 10^{-6}, 10^{-7}$. 

On the other hand, one can derive the upper limits of spin-down plateau luminosity ($L_{0}$) for varying $\eta_{\rm R}=10^{-4}, 10^{-5}, 10^{-6}, 10^{-7}$. This means that the observed X-ray plateau emission should exist before the coherent radio emission. The left panel of Figure 6 shows the upper limit of pseudo-plateau luminosity as a function of time in the rest frame and then compares with the observed X-ray data from Swift/XRT. It is clear to see that no X-ray plateau emission is observed below $L_{0}$ for $\eta_{\rm R}=10^{-4}$ and $10^{-5}$, but it may exist for $\eta_{\rm R}=10^{-6}$ and $10^{-7}$, due to insufficient observational data before and after the collapse time. It also supports a preferably lower value of $\eta_{\rm R}$.

Moreover, the collapse time of GRB 201006A is the longest one by comparing with other short GRBs if its central engine is a supramassive magnetar, and it is another reason to make GRB 201006A as an interesting case. In the right panel of Figure 6, we also compare the $L_{X}-t_{\rm col}$ correlation of GRB 201006A with that of other short GRBs. We find that it also follows the same anticorrelation as that of other short GRBs, and it also supports that they share the same physical process.

\subsection{Possible kilonova emission}
The mergers of binary NSs are typically accompanied by an abundance of electromagnetic transients, such as short GRBs and their afterglow emissions \citep{2011ApJ...732L...6R, 2016ApJ...827..102T}, and an optical/infrared transient (called a kilonova) powered by radioactive decay from r-process \citep{1998ApJ...507L..59L, 2010MNRAS.406.2650M, 2011ApJ...732L...6R, 2013PhRvD..87b4001H, 2013MNRAS.430.1061R}. If the central engine of GRB 201006A is a supramassive magnetar, then the main power of the possible kilonova emission should no longer be limited to the r-process, but the spin energy of the magnetar should also be considered \citep{2013ApJ...776L..40Y, 2014MNRAS.439.3916M, 2017ApJ...837...50G, 2021ApJ...912...14Y, 2022MNRAS.516.4949S, 2024MNRAS.527.5166W, 2024arXiv240500638A}.

From an observational point of view, several optical telescopes are following up to observe the GRB 201006A, such as GROWTH-India Telescope \citep{2020GCN.28573....1K}, MITSuME Akeno \citep{2020GCN.28571....1I} and Lowell Discovery Telescope \citep{2020GCN.28572....1D}, but they did not find any new point sources within the enhanced Swift/XRT circle, excepting the upper limits. In order to test whether the upper limits of optical observations can be used to constrain the possible kilonova emission associated with GRB 201006A, we numerically calculate the light curve of kilonova emission in $i$ and $r$ bands. In our calculations, we adopt the ejecta mass in the range of $M_{\rm ej}=(10^{-4}-10^{-2})\rm~ M_{\odot}$, which are typical values from numerical simulation of NS–NS mergers \citep{2013PhRvD..87b4001H}, a relativistic speed $\beta = (0.1-0.3)$, and opacity $\kappa = (0.1-10) ~\rm cm^{2}~g^{-1}$. The total energy of the ejecta and shocked medium can be expressed as $E=(\Gamma -1)M_{\rm ej}c^2 + \Gamma E_{\rm int}^{'} $, where $\Gamma$ and $E_{\rm int}^{'}$ are the Lorentz factor and the internal energy in the comoving frame, respectively. Based on the energy conservation, $d E_{\rm ej} = (L_{\rm inj} - L_{\rm e})dt$, where $L_{\rm e}$ and $L_{\rm inj}$ are the radiated bolometric luminosity and injection luminosity, respectively. $L_{\rm inj}$ is contributed by spin-down luminosity ($L_{\rm 0}$) from the magnetar and radioactive power ($L_{\rm ra}$), namely, $L_{\rm inj}=\xi L_{\rm 0} + L_{\rm ra}$. Here, we adopt $\xi=0.1$, which is converted efficiency from spin-down luminosity to thermal energy of the ejecta \citep{2011ApJ...726...90Z}. The equations of the full dynamic evolution of the ejecta, the evolution of internal energy in the comoving frame, and the evolution of the comoving volume can be found in \cite{2013ApJ...776L..40Y} and \cite{2021ApJ...912...14Y}. For the spin-down energy of the magnetar, we adopt a timescale of energy injection $t=4600$ s and different initial luminosities of energy injection, e.g., $L_{0} = 2.76\times10^{46}~ \rm erg~s^{-1}$, $L_{0} = 2.76\times10^{47}~ \rm erg~s^{-1}$, and $L_{0} = 2.76\times10^{48}~ \rm erg~s^{-1}$, respectively.

Figure 7 shows the light curves of kilonova emission in $i$ and $r$ bands with different $L_{0}$, and we find that the expected kilonova emission is much fainter than the upper limit of observed optical data. This means that the possible kilonova emission associated with GRB 201006A is too faint to be detected. Hence, the possibility that the central engine is a supramassive NS cannot be ruled out, which is inconsistent with the results in  \cite{2024ApJ...973L..20S}.

\section{Conclusion and Discussion}
GRB 201006A is a short-duration burst with $T_{90} \sim 1.7$ s in 50-300 keV and was detected by both Swift and Fermi. The $\gamma-$ray light curve consists of a single hard-spike. By extracting the spectra of GRB 201006A, we find that the CPL model is the best-fitting model and gives a soft $E_{\rm p} = 103\pm58~\rm keV$. Recently, \cite{2024MNRAS.tmp.2177R} claimed the detection of a coherent radio flash at 76.6 minutes after the burst trigger time of GRB 201006A, and one has to estimate the redshift ($z=0.58\pm0.06$) of GRB 201006A via dispersion of radio emission. Hence, the isotropic energy at redshift $z = 0.58$ is estimated as $E_{\rm \gamma,iso} = 2.76^{+0.51}_{-0.47} \times 10^{50} \rm erg$. By calculating $\varepsilon \sim 0.012$ and $E_{\rm \gamma,iso}-E_{\rm p}$, we suggest that the progenitor of GRB 201006A is likely to be from merger of compact stars.

However, the physics mechanism behind this coherent radio flash is still under debate. \cite{2024MNRAS.tmp.2177R} proposed that the coherent radio flash is powered by the collapse of a long-lasting supramassive magnetar as the central engine into a BH. \cite{2024ApJ...973L..20S} suggested that the central engine is most likely a BH, and the coherent radio flash is produced in regions far from the BH through mechanisms such as a synchrotron maser or magnetic reconnection.

In this paper, we first attempt to test whether the radio flash is caused by an early radio afterglow produced by the synchrotron radiation from the interaction between the jet and environmental medium. It is found that an early radio afterglow as the physical origin of the radio flash can be ruled out. Then, one considers the hypothesis of a supramassive magnetar which is the central engine before collapsing into a BH \citep{2024MNRAS.tmp.2177R}. We find that the derived magnetar surface magnetic field ($B_{\rm p}$) and the initial spin period ($P_{\rm 0}$) fall into a reasonable range but require a preferably low value of $\eta_{\rm R} = 10^{-7}$ or $10^{-6}$. Moreover, we numerically calculate the possible kilonova emission in the $i$ band and $r$ band by considering both the r-process and energy injection from the magnetar. It is found that the calculated luminosity of a kilonova is below the upper limits of optical observations, and it may be too faint to be detected. No detection associated with kilonova emission with GRB 201006A is not contradictory to that of the supramassive magnetar as the central engine that survived for 49 minutes.

\cite{2024ApJ...973L..20S} dismissed the perspective that the central engine is a millisecond magnetar owing to the absence of kilonova observations. However, it is worth noting that the simulated ejecta in their model exhibits a higher mass by comparing with that of binary NS mergers in numerical simulations, e.g., $M_{\rm ej}=(10^{-4}-10^{-2})\rm~ M_{\odot}$ \citep{2013PhRvD..87b4001H}. Hence, we believe that a supramassive magnetar remains a potential candidate of the central engine for GRB 201006A, which is consistent with the observations.

\section{Acknowledgments}
We acknowledge the use of the public data from the Swift and Fermi Science Data Center. This work is supported by the Natural Science Foundation of Guangxi (grant No. 2023GXNSFDA026007), the Natural Science Foundation of China (grant Nos. 11922301 and 12133003), and the Program of Bagui Scholars Program (LHJ).

{}

\clearpage


\begin{table*}[h]\footnotesize %
 \centering
  \caption{The derived parameters or upper limits of the magnetar for different efficiency $\eta_{R}$.}
  \setlength{\tabcolsep}{2mm}{
  \begin{center}
  \renewcommand\arraystretch{1.5}
  \begin{tabular}{cccccccc}
  \hline
  \hline
  $\eta_{\rm R}$ & $P_{0}$ & $B_{\rm p,15}$ & $L_{0}$ \\
        &   (ms)  &  (G)       &   ($\rm erg~s^{-1}$)\\
  \hline
  $10^{-7}$&	5.06$\uparrow$&	4.25& $2.76\times10^{47}\downarrow$\\
  $10^{-6}$&	1.60$\uparrow$&	1.34& $2.76\times10^{48}\downarrow$\\
  $10^{-5}$&	0.51$\uparrow$&	0.42& $2.76\times10^{49}\downarrow$\\
  $10^{-4}$&	0.16$\uparrow$&	0.13& $2.76\times10^{50}\downarrow$\\
  \hline
\end{tabular}
\end{center}}
\end{table*}



\begin{figure*}[htbp!]\label{fig:lightcurve}
\center
\includegraphics[angle=0,width=0.5\textwidth]{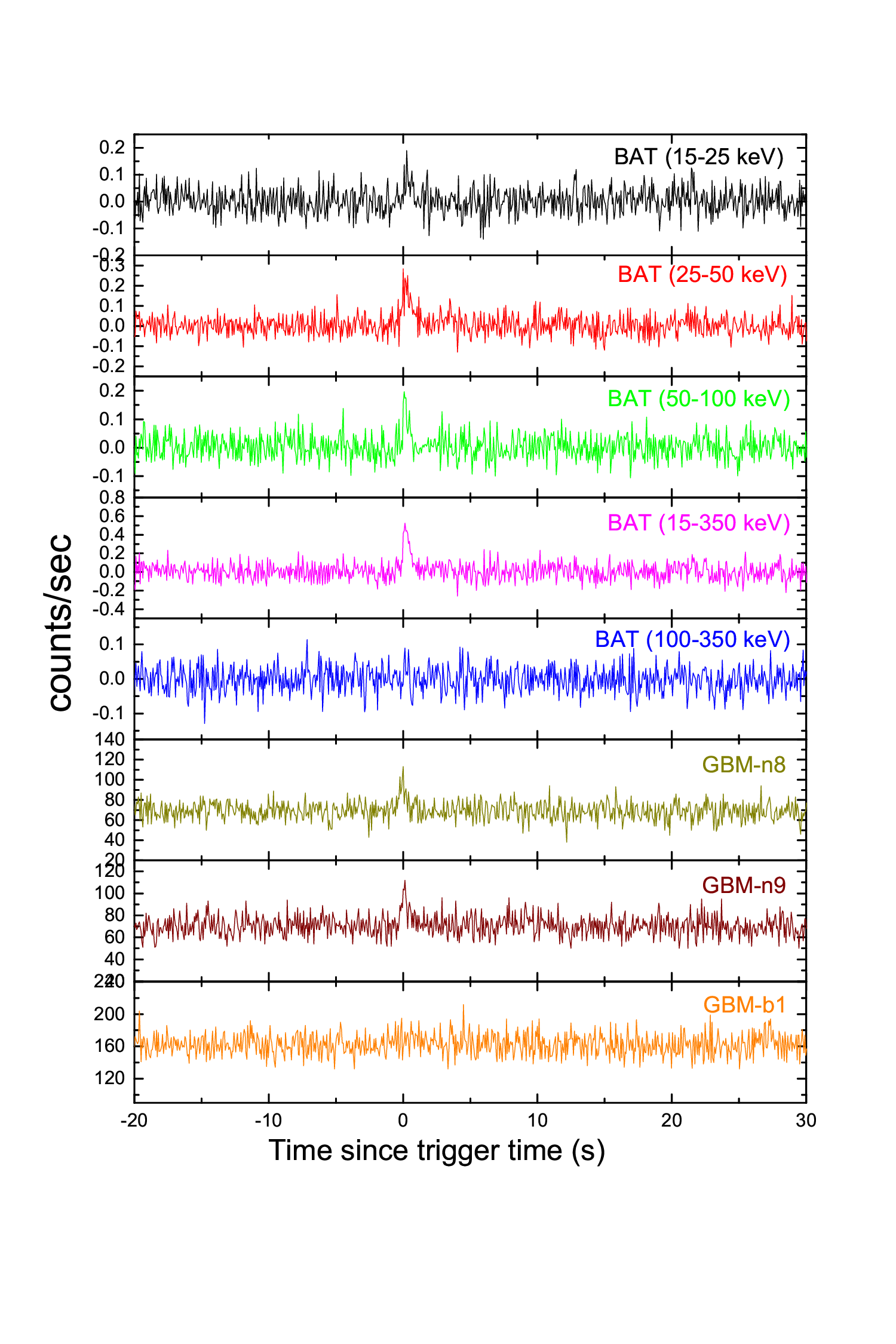}
\caption{Swift/BAT and Fermi/GBM light curves of GRB 201006A in different energy bands with a 128 ms time bin.}
\end{figure*}

\begin{figure*}[htbp!]\label{fig:Energy spectrum}
\center
\rotatebox[origin=c]{270}{\includegraphics[width=0.3\textwidth]{f2a.eps}}
\includegraphics[angle=0,width=0.4\textwidth]{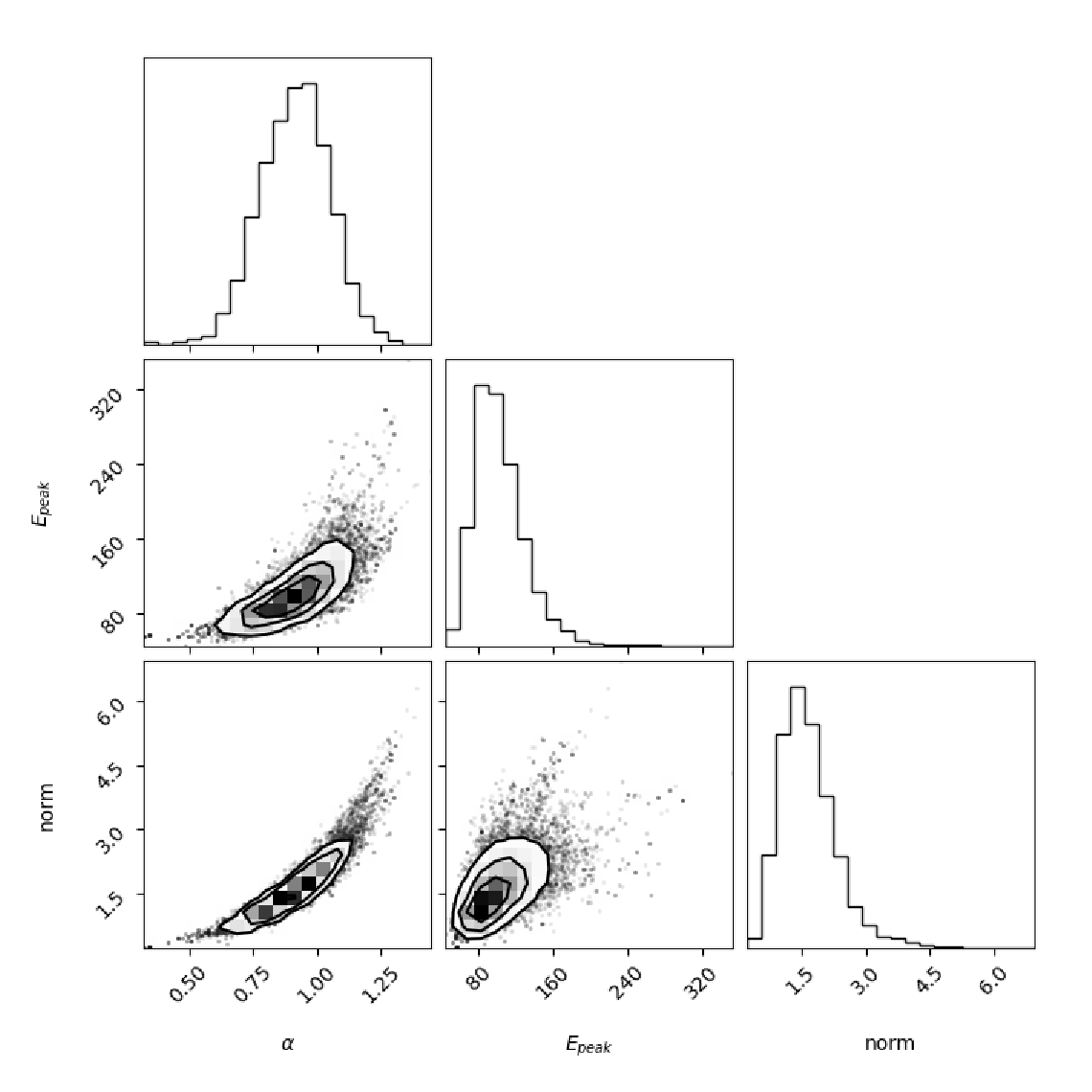}
\caption{Spectral fit results of GRB 201006A from $T_{0}$-0.19~s to $T_{0}$+0.83~s with the CPL model for Fermi/GBM, including the photon spectrum (left) and parameter constraints (right) of the CPL fit.}
\end{figure*}

\begin{figure*}[htbp!]\label{fig:T90 Ep-Eiso }
\center
\includegraphics[angle=0,width=0.45\textwidth]{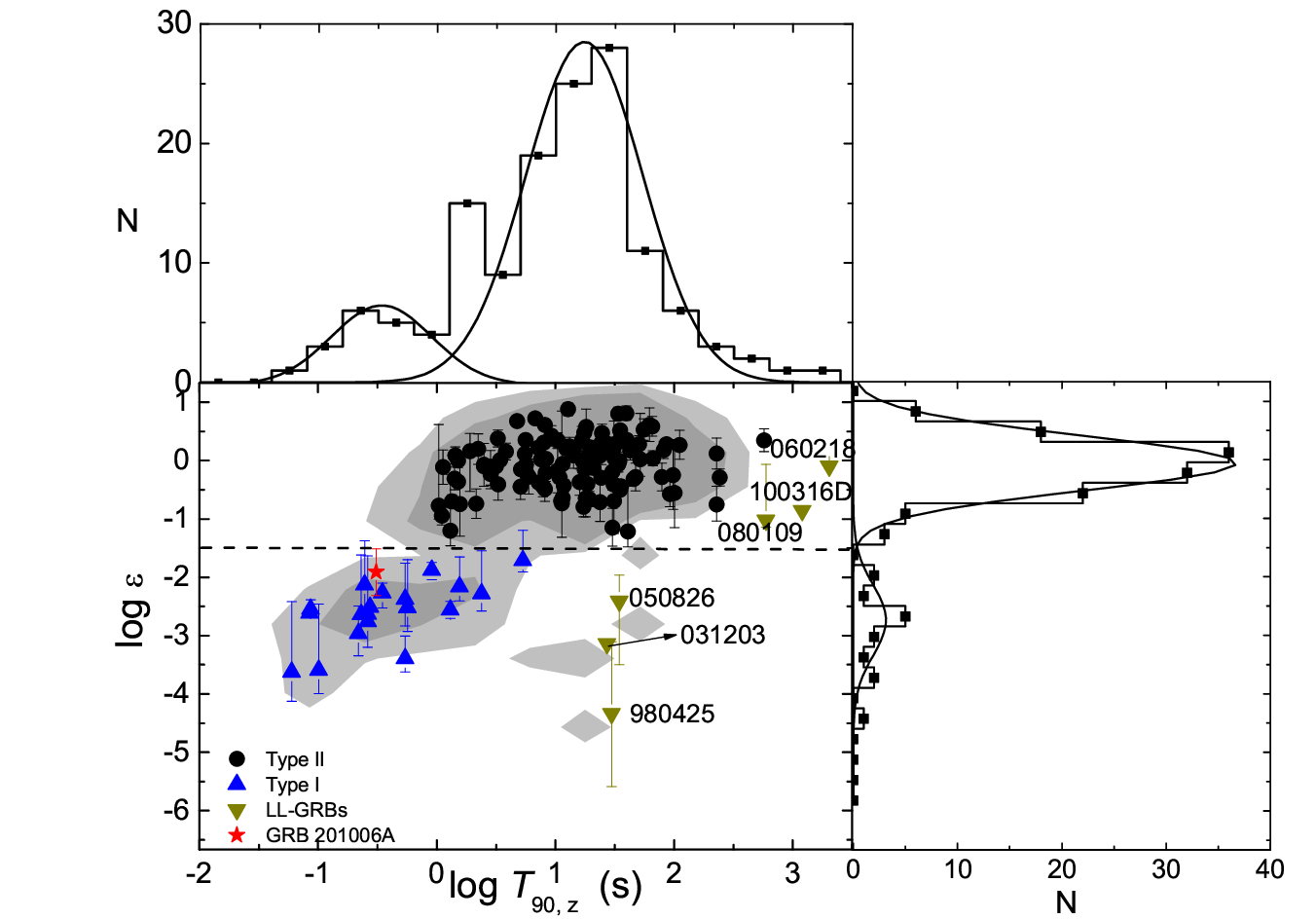}
\includegraphics[angle=0,width=0.5\textwidth]{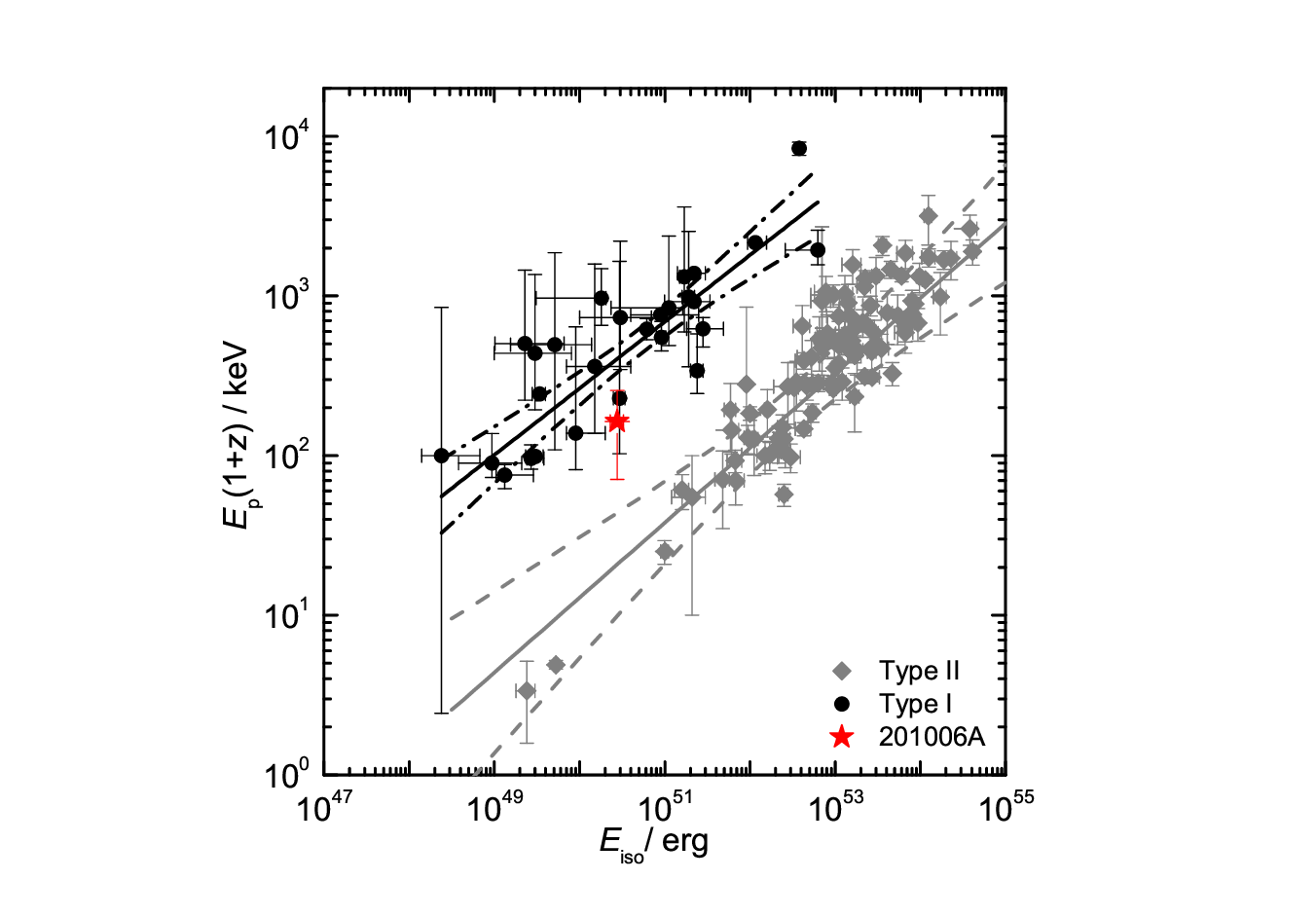}
\caption{Left: 1D and 2D distributions of GRBs samples in $T_{90}-\epsilon$ space. The dashed line indicates $\epsilon = 0.03$, and the data come from \cite{2010ApJ...725.1965L}. Right: $E_{\rm p}-E_{\rm \gamma,iso}$ correlation diagram. Gray diamonds and black circles correspond to type II and type I GRBs, respectively, which are from \cite{2002A&A...390...81A} and \cite{2009ApJ...703.1696Z}. The red star indicates GRB 201006A. The black and gray solid lines represent the best-fit lines, and dashed lines represent the 3$\sigma$ confidence bands.}
\end{figure*}

\begin{figure*}[htbp!]\label{fig:FS RS}
\center
\includegraphics[angle=0,width=0.45\textwidth]{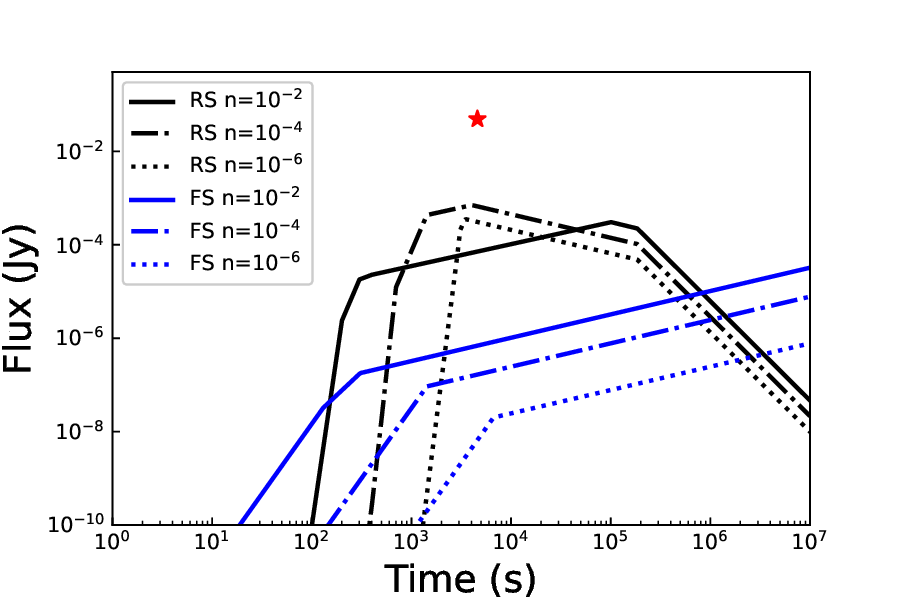}
\includegraphics[angle=0,width=0.45\textwidth]{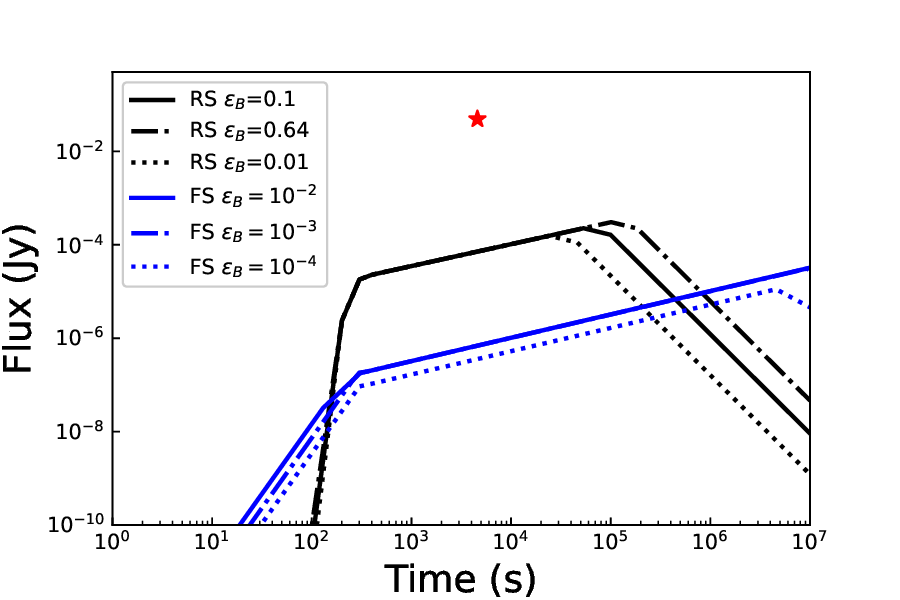}
\caption{Numerical calculation of FS and RS afterglow light curves in radio band of GRB 201006A. The model parameters are $\epsilon_{\rm e} = 0.1$, $p = 2.5$ and $\Gamma = 100$. Left: the case of FS (blue) and RS (black) light curves at different circumburst medium densities. Different shock microphysical parameters $\epsilon_{\rm B}$ are adopted: $\epsilon_{\rm B,f} = 0.01$ for the FS model and $\epsilon_{\rm B,r} = 0.64$ for the RS model. Right: the case of FS and RS light curves under different shock microphysics parameters $\epsilon_{\rm B}$. We apply the typical circumburst density value ($n = 10^{-2}$~$\rm cm^{-3}$) of SGRBs to calculate the afterglow flux. The red star represents the observed peak flux density $49\pm27$ mJy of the radio flash with a redshift of $0.58 \pm 0.06$ \citep{2024MNRAS.tmp.2177R}.
}
\end{figure*}

\begin{figure*}[htbp!]\label{fig:P0-Bp}
\center
\includegraphics[angle=0,width=0.5\textwidth]{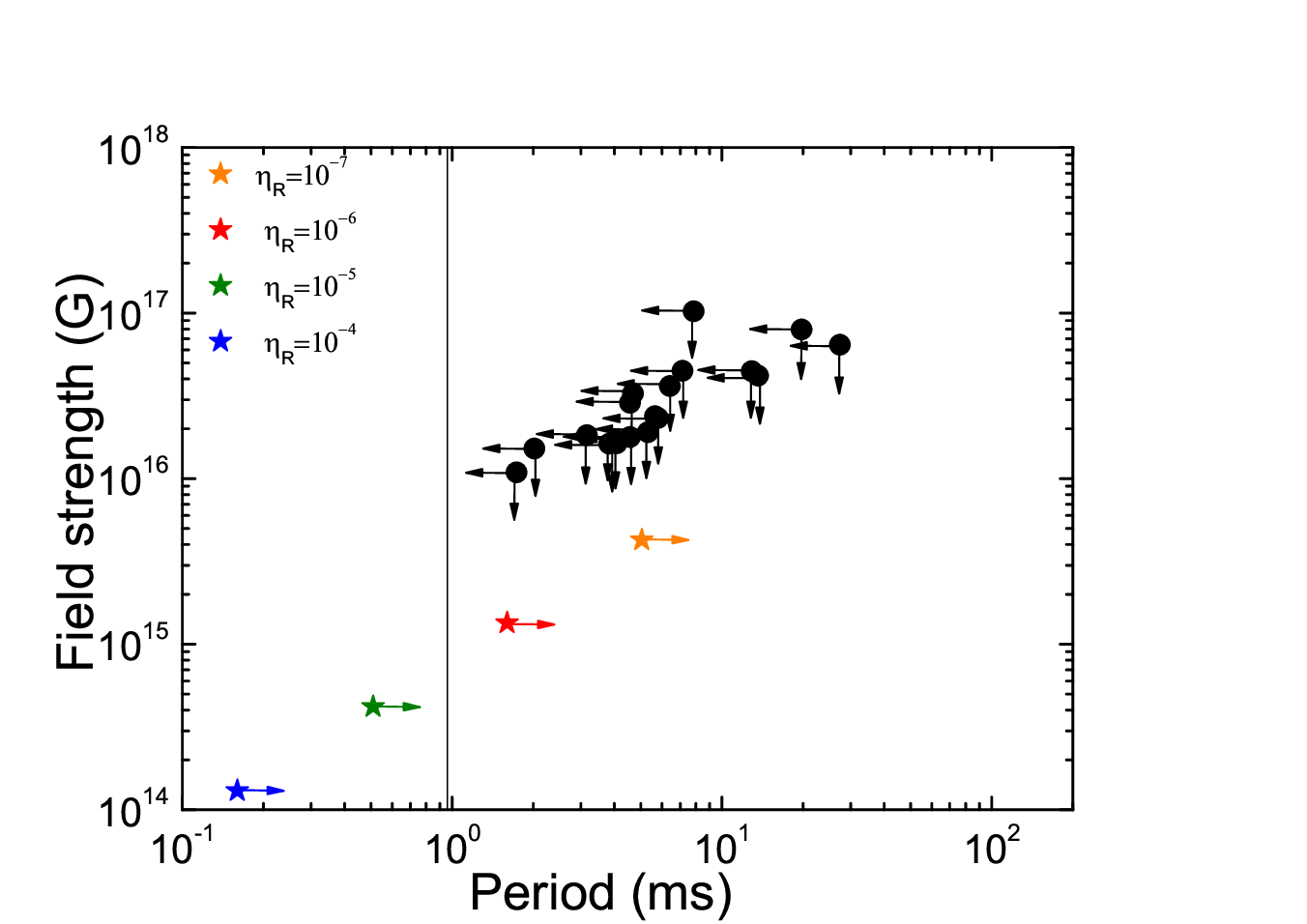}
\caption{Inferred magnetar parameters. Initial spin period $P_{0}$ vs. surface polar cap magnetic field strength $B_{p}$ derived by short GRBs with internal plateaus (black circles) from \cite{2015ApJ...805...89L}. In the case of $\eta_{\rm R}=10^{-4}, 10^{-5}, 10^{-6}$, and $10^{-7}$ of GRB 201006A are maked as stars. The vertical solid line is the breakup spin period limit for an NS \citep{2004Sci...304..536L}.}
\end{figure*}

\begin{figure*}[htbp!]\label{fig:L0 Lx}
\center
\includegraphics[angle=0,width=0.45\textwidth]{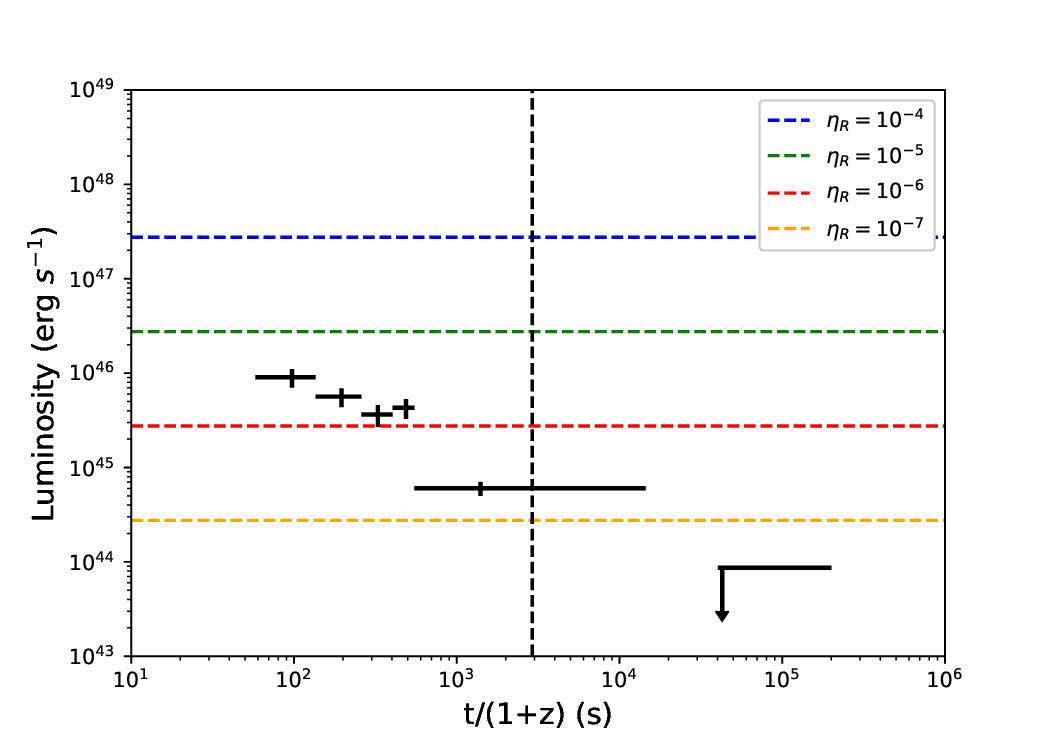}
\includegraphics[angle=0,width=0.45\textwidth]{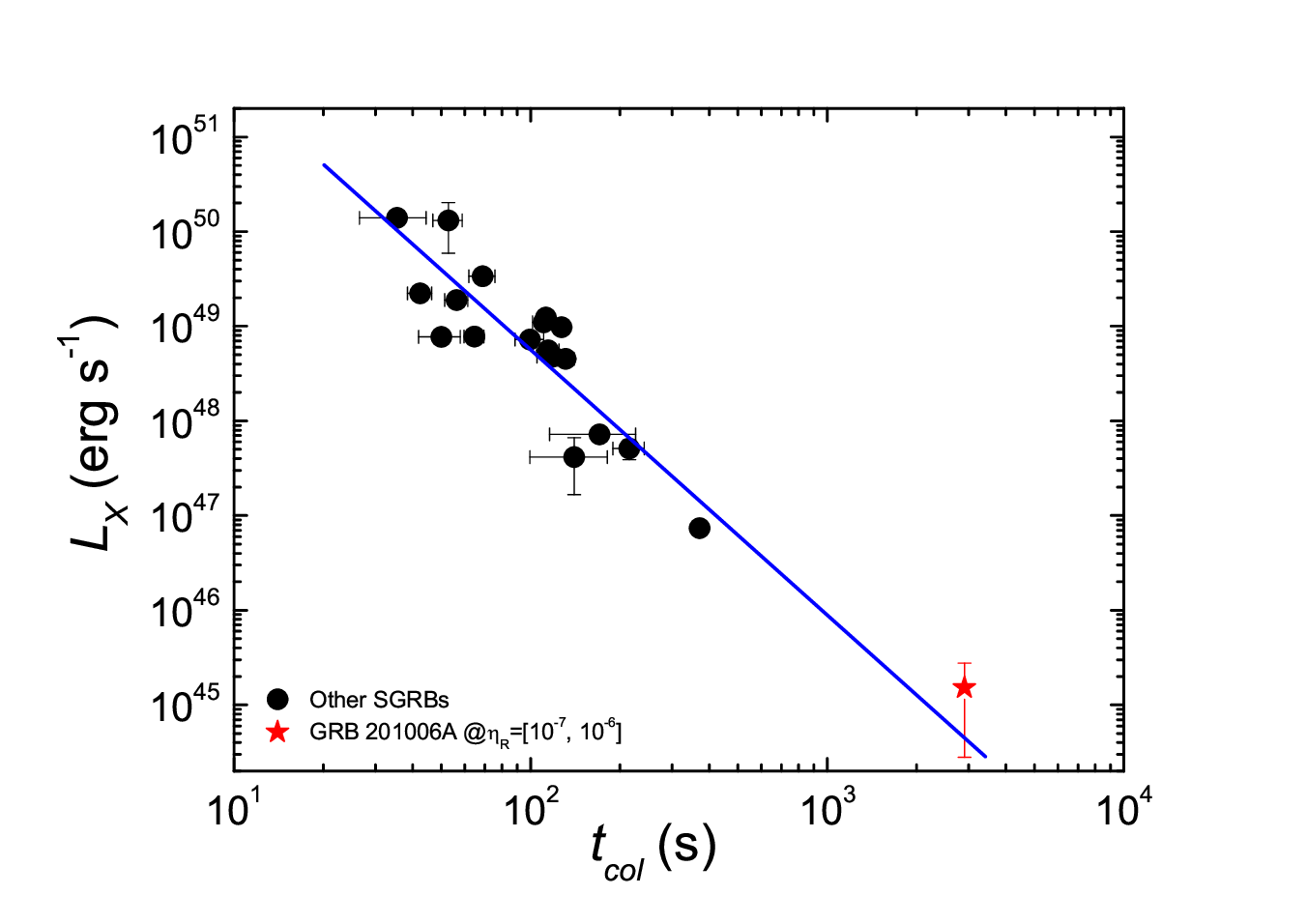}
\caption{Left: the inferred X-ray plateau luminosity at various radiation efficiencies $\eta_{\rm R}$. The data points are taken from the swift website, and the black dashed line represents the burst time of the radio flash (in rest frame). Right: $L_{\rm X}-t_{\rm col}$ anticorrelation for short GRBs with internal plateaus. The black circles indicate the samples from  \cite{2015ApJ...805...89L}, and the red star is GRB 201006A.}
\end{figure*}
\begin{figure*}[htbp!]\label{fig:kilonova}
\center
\includegraphics[angle=0,width=0.4\textwidth]{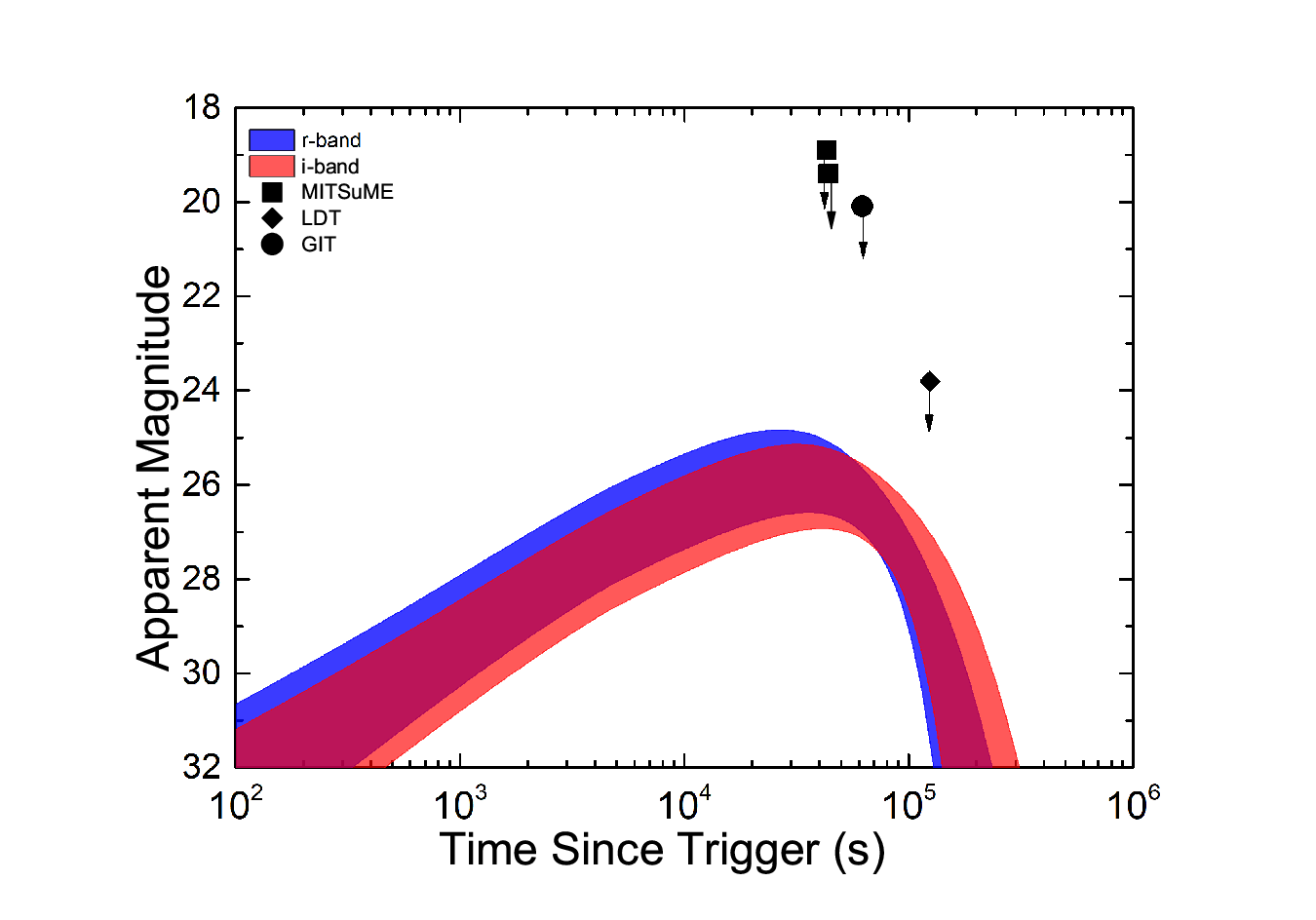}
\includegraphics[angle=0,width=0.4\textwidth]{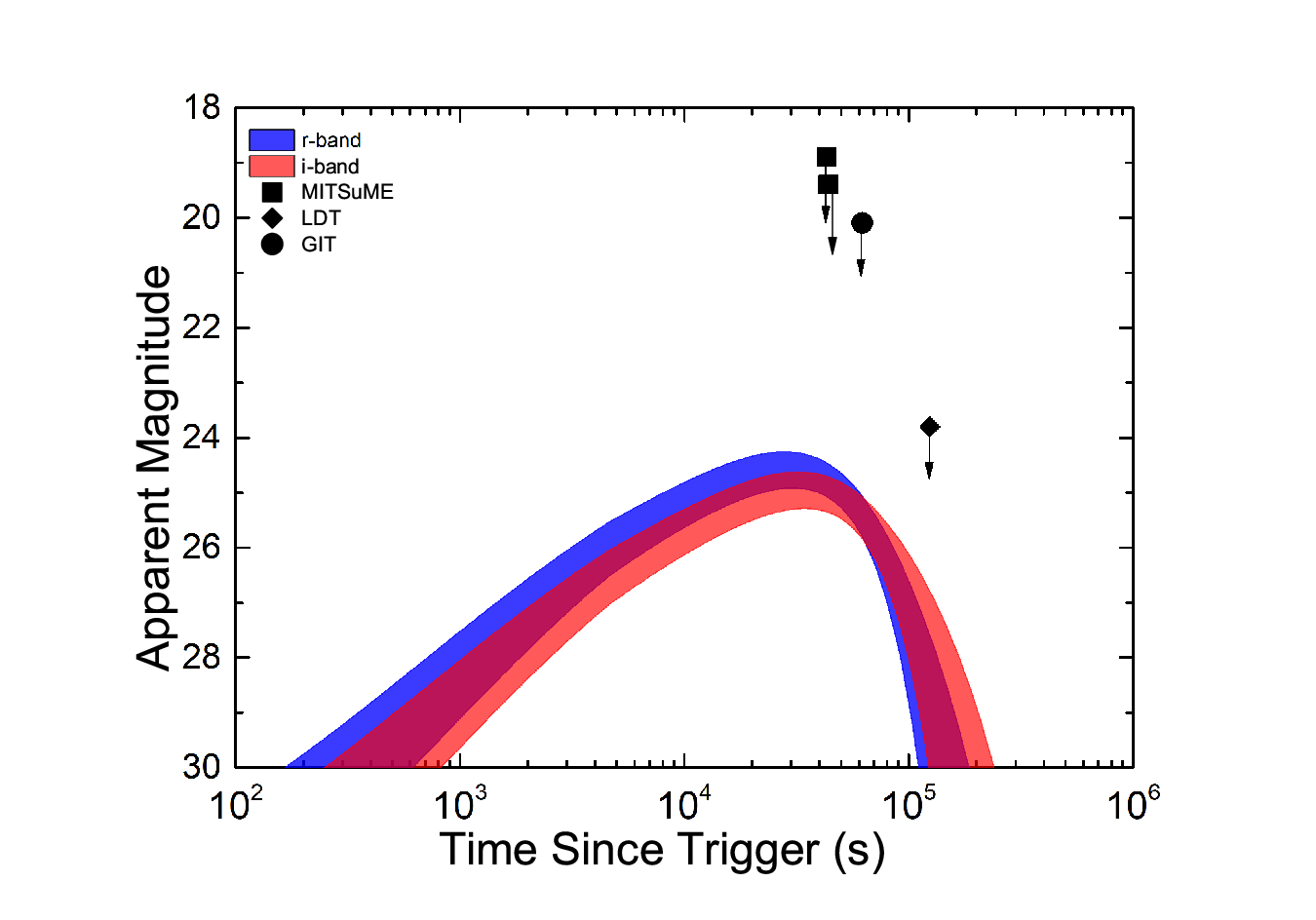}
\includegraphics[angle=0,width=0.4\textwidth]{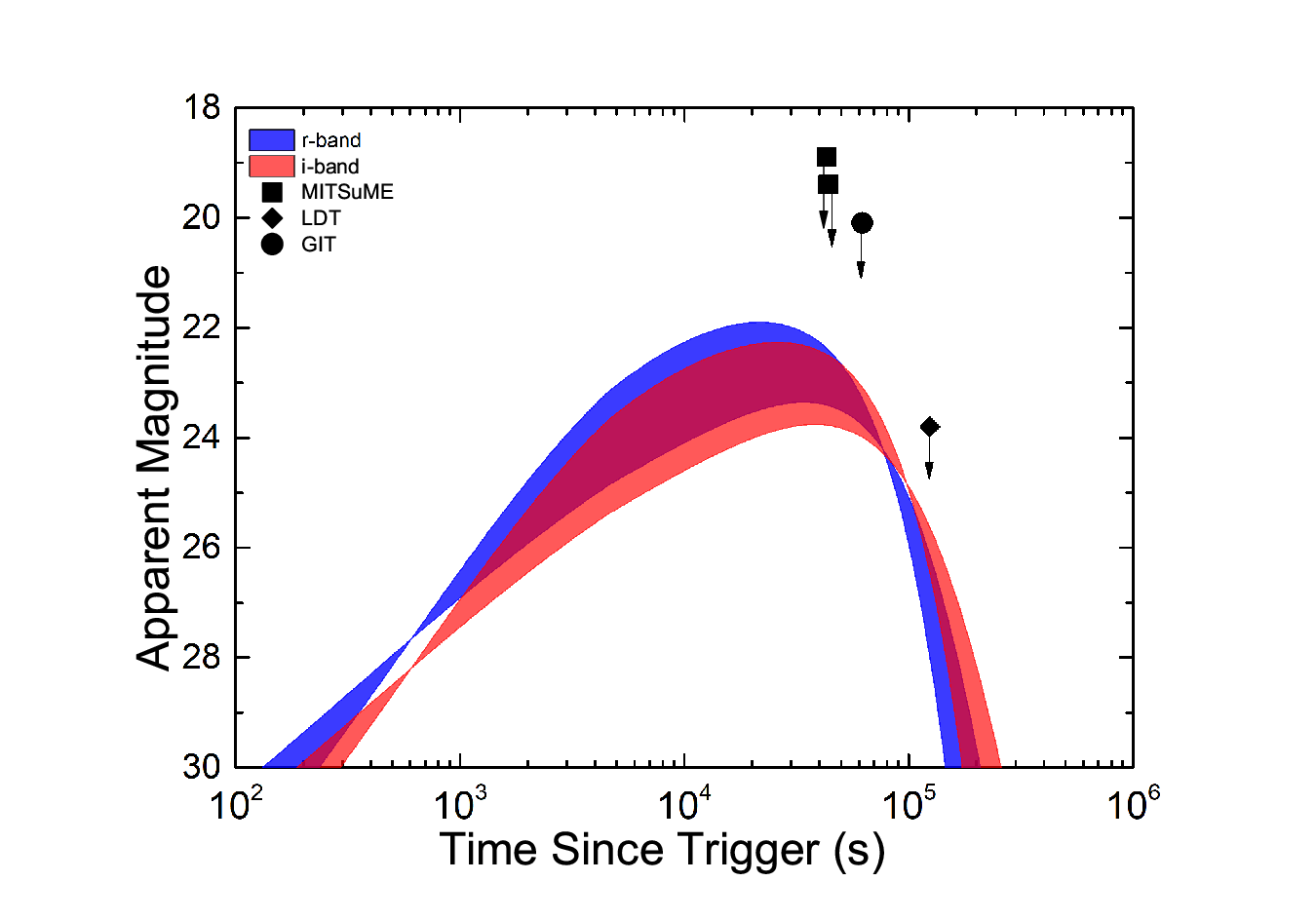}
\caption{Magnetar-driven kilonova light curves in $i$ band (red) and $r$ band (blue). From left to right: the kilonova light curves under three different initial injection luminosity, $L_{0} = 2.76\times10^{46}~ \rm erg~s^{-1}$, $L_{0} = 2.76\times10^{47}~ \rm erg~s^{-1}$, and $L_{0} = 2.76\times10^{48}~ \rm erg~s^{-1}$, respectively. The other parameters are in the ranges of $M_{\rm ej} = 10^{-4}$ to $10^{-2}~\rm M_{\odot}$, $\beta = 0.1-0.3$, and $\kappa = 0.1-10 ~\rm cm^{2}~g^{-1}$. The black square, circle and diamond correspond to the upper limits of optical observations.}
\end{figure*}
\end{document}